\documentclass[conference]{IEEEtran}
\IEEEoverridecommandlockouts
\usepackage{cite}
\usepackage{amsmath,amssymb,amsfonts}
\usepackage{algorithmic}
\usepackage{graphicx}
\usepackage{textcomp}
\usepackage{xcolor}
\usepackage{tikz}
\usepackage{subfigure}
\usepackage{bm}
\usepackage{amsmath}
\usepackage{multirow}
\usepackage{balance}
\usetikzlibrary{fit,positioning}
\usepackage{enumitem}
\def\BibTeX{{\rm B\kern-.05em{\sc i\kern-.025em b}\kern-.08em
    T\kern-.1667em\lower.7ex\hbox{E}\kern-.125emX}}
\begin{document}

\title{Scientific Dataset Discovery via Topic-level Recommendation
}

\author{\IEEEauthorblockN{Basmah Altaf}
\IEEEauthorblockA{CEMSE \\
\textit{King Abdullah University}\\ \textit{of Science and Technology}\\
basmah.altaf@kaust.edu.sa}
\and
\IEEEauthorblockN{Shichao Pei}
\IEEEauthorblockA{CEMSE \\
\textit{King Abdullah University}\\ \textit{of Science and Technology}\\
shichao.pei@kaust.edu.sa}
\and
\IEEEauthorblockN{Xiangliang Zhang}
\IEEEauthorblockA{CEMSE \\
\textit{King Abdullah University}\\ \textit{of Science and Technology}\\
xiangliang.zhang@kaust.edu.sa}
}

\maketitle

\begin{abstract}
Data intensive research requires the support of appropriate datasets. However, it is often time-consuming to discover usable datasets matching a specific research topic. We formulate the dataset discovery problem on an attributed heterogeneous graph, which is composed of paper-paper citation, paper-dataset citation and also paper content. We propose to characterize both paper and dataset nodes by their commonly shared  latent topics, rather than  learning user and item representations via canonical graph embedding models, 
because the usage of datasets and the themes of research projects can be understood on the common base of research topics. The relevant datasets to a given research project can then be inferred  in the shared topic space.
The experimental results show that our model can generate reasonable profiles for datasets, and recommend proper datasets for a query, which represents  a research project linked with several papers.
\end{abstract}


\section{Introduction}
Accessing to datasets is crucial for researchers and scientists to evaluate their new scientific discoveries, and reproduce research results \cite{brickley2019google}. Dataset repositories provide a common way to access millions of datasets of different fields ranging from scientific data, commercial data to government data and more. 
Despite the availability of plentiful online dataset repositories, it is not easy for users to locate datasets that match their research interests.  
First, the number of datasets that appear in dataset search are continuously growing.
Some well-known repositories (e.g., Kaggle \cite{kaggle}, re3data \cite{rueda2017datacite}, DataCite \cite{kindling2017landscape}, Linked Open Data Cloud \cite{LOD}) 
are not comprehensive as many new datasets are being collected and published
that may be unknown to users. 
Second, 
the existing methods allow researchers to find scientific datasets by (a) reading a mass of relevant papers or (b) using search engines such as Google Dataset search\footnote{https://toolbox.google.com/datasetsearch}. 
However, the former method is a time-consuming process, and the latter method usually 
requires to refine the query words {repeatedly} after checking the search results, which may be difficult for entry-level researchers  due to their shallow understanding about their research problem \cite{altaf2019dataset}. Both these methods fail to consider the common semantics between user's research interests and available datasets.

\begin{figure}[!t] 
	\centering
	\includegraphics[scale=0.30]{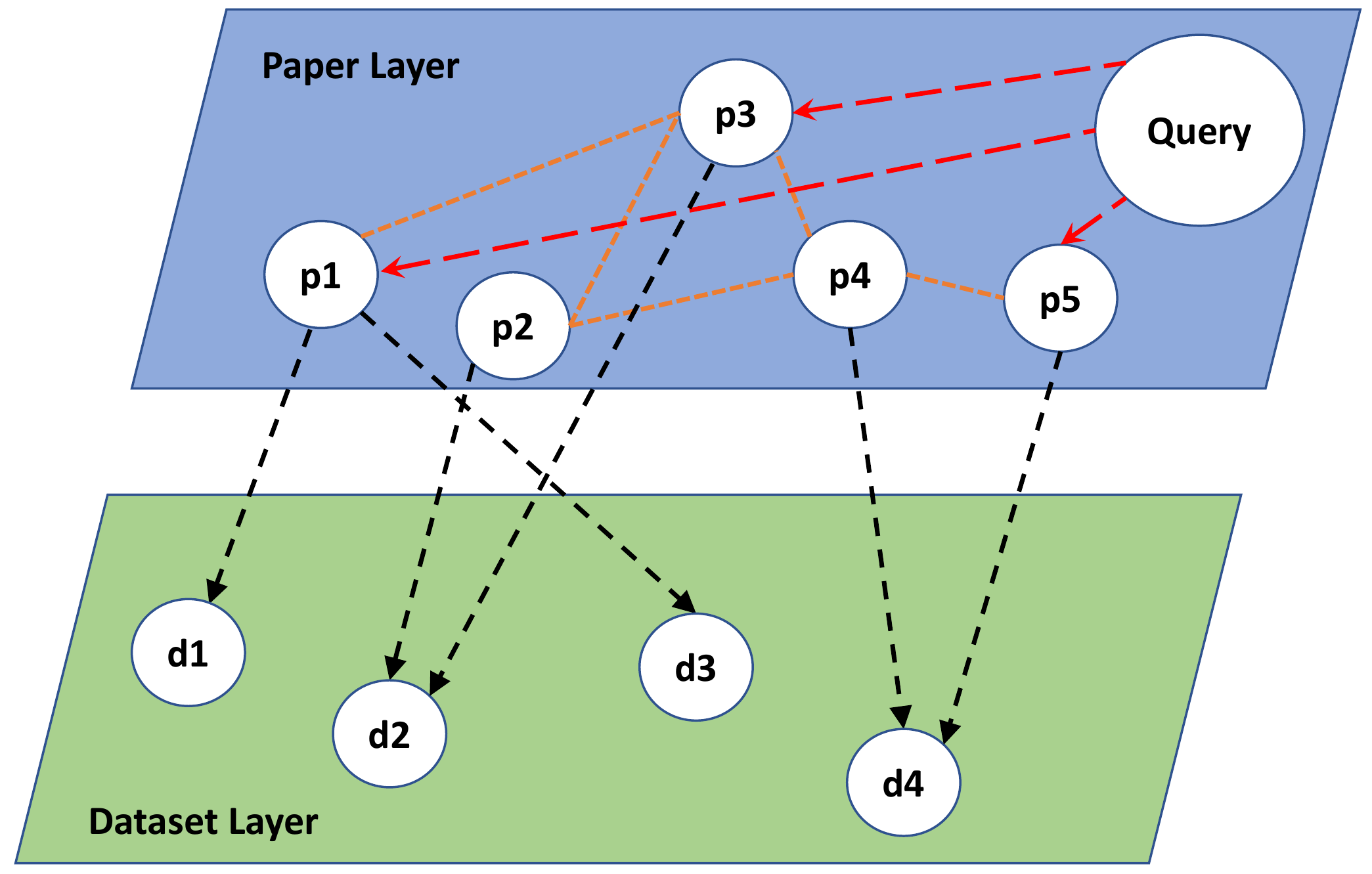}
	\caption{Illustration of query-based dataset recommendation problem. Given a paper-dataset heterogeneous network with paper content, our objective is to recommend datasets to answer a query $Q$, which indicates interest in paper p1, p3 and p5.} 
	\label{fig:paperDatasetGraph}
\end{figure}

Inspired by the recent progress of recommendation system, in this paper, we formulate the scientific datasets discovery as a recommendation task.
Given an attributed heterogeneous network that is composed of paper-paper citation, paper-dataset citation and paper content, as Figure \ref{fig:paperDatasetGraph} illustrated, our task is to recommend datasets to answer a query $Q$, which for example is a research project being relevant to paper p1, p3 and p5. A naive and obvious answer is to return datasets d1, d2, d3 and d4, because they are directly linked with query papers. However, there always exist other usable datasets that are relevant at topic level to the query papers, but not directly linked.
We have to identify common topics shared by paper and dataset nodes in order to recommend datasets to  researchers based on their research interests indicted in the query.
Motivated by the usefulness of topic modeling, 
we aim at designing a topic-modeling based dataset recommendation approach \textit{that can infer the common latent topics of both the paper and dataset nodes}.

Although recommendation on heterogeneous graphs have been widely studied,
the need of reasonable profiles for items and users in some industrial applications cannot be satisfied.
With the surge of topic modeling techniques \cite{blei2003latent}, researchers find that the probabilistic topic modeling can largely improve the content-based recommendation  \cite{pazzani2007content}  
which takes the content information (e.g., reviews, attributes) into consideration by learning the distribution over terms in text to provide an interpretable low-dimensional representation.
Nevertheless, the existing topic modeling based recommendation approaches have the following limitations.  
(1) \textbf{Non-shared topic distributions}. There are topic models applied to heterogeneous graphs \cite{ma2009learning, ma2008sorec,Fan:2019:GNN,wu2019dual} and attributed  heterogeneous graphs \cite{chen2014context,  hu2015synthetic, hu2018collaborative,ma2011recommender}. These all approaches profile users and items by different latent topic distributions.  Learning user and item topic distribution in separated spaces is reasonable when they are from different domains (e.g., users and items are not from the same domain). However, in scenarios like paper recommendation, citation recommendation, and other scholarly recommendation, nodes (e.g., papers, authors, venues) although of different types, share common topics. It is thus necessary to infer the shared latent topics for all nodes. 
(2) \textbf{Ignorance of structure and text modeling of attributed network.} Current topic modeling based recommendation solutions only  model the statistical co-occurrences of words in text (e.g., review/comments), without considering the global representation of text. Nevertheless, global representation has been demonstrated \cite{bansal2016ask, kim2016convolutional, seo2017interpretable} to be essential and effective for text modeling. Moreover,  
recent network embedding models \cite{perozzi2014deepwalk, tang2015line} demonstrate that the structure information of a network plays a significant role in improving the performance of relation discrimination. Therefore, it is worth  exploring the joint modeling of both  text and structure information of nodes with common topics.  

The aforementioned objective and the limitations of current solutions motivated us to propose a method \textbf{STopic-Rec} (recommendation based on shared topic distribution), 
which contains a probabilistic topic modeling module and a deep learning based joint representation learning module. 
Specifically, we first design a probabilistic topic modeling module to learn word (of paper) distributions and dataset distributions over shared topics. It is natural to use a shared topic distribution to couple the datasets and the text 
because the text of a paper and the used datasets in the paper usually share the same theme. Meanwhile, the learned topic distribution over datasets can be utilized to generate the dataset profile for search engines.
Then we develop a joint representation learning module to learn the paper's embedding  by encoding both (a) the global information of text and (b) structure preserving information of nodes in a network. In particular, we leverage the Stacked Denoising Autoencoder \cite{vincent2010stacked} to model the global semantic information of text, and a network embedding method to preserve network neighborhoods of nodes. 
Next, inspired by the HFT model\cite{mcauley2013hidden}, we design a transformation function to fuse the learned topics distributions and learned embedding vectors for papers, in order to inject the joint information of text and heterogeneous graph structure into the topic generation process. Finally, we jointly optimize the probabilistic topic modeling module along with joint representation learning module using alternating gradient descent.

Our contributions in this work are summarized as follows:
\begin{itemize}
	\item We exploit the \textbf{shared topic} information of nodes in attributed heterogeneous graph to improve the performance of query-based dataset recommendation, meanwhile also generating \textbf{topic profiles} for each dataset.
	
	
	\item 
	We also design a joint representation learning module to encode node text and node structural property (citation graph). Moreover, a transformation function is employed to discover topics that are correlated with the hidden factors of nodes considering both the text and structure information.
	
	\item We conduct extensive experiments on a real-world dataset and show that our proposed approach outperforms the state-of-the-art methods for dataset recommendation task.
\end{itemize}

\section{Related Work}\label{sec:relatedwork}
\subsection{Social Recommendation}
Social recommendation utilizes social information as the assistance to alleviate data sparsity and cold-start problems in traditional collaborative filtering based recommendation systems \cite{tang2013social}.
Some matrix factorization based approaches propose to factorize user-item rating matrix and user-user linkage matrix simultaneously \cite{ma2009learning}, or aggregate a user's rating and his friends' ratings to predict the target user's rating on an item \cite{ma2008sorec}. Recent works \cite{Fan:2019:GNN,wu2019dual} employed graph convolutional networks to address the multi-hop problem for learning from both user-user and user-item matrices. 

Our problem is analogous to incorporating social relations in recommendation. In our scenario, papers are equivalent to users and datasets are analogous to items. Moreover, the paper citation network can be regarded as the social relationship among users.
However, current solutions for social recommendation only consider the social relations between nodes for recommendation problem while ignoring the node attributes. 
In our work, we learn semantics of papers from their associated rich text information and also associated text information of their neighboring nodes in order to leverage the heterogeneous information of social relations.

\subsection{Query based Recommendation}
Query based recommendation aims to recommend a list of items to the target user according to the previous preference of the user and an input query \cite{chen2016query}. For instance, a query-based music recommendation \cite{chen2016query} via preference embedding   allows a user to explore new song tracks by specifying either a track, an album, or an artist. Furthermore, citation recommendation methods accept a given manuscript as the query, and return a list of reference papers that can be cited by this new draft \cite{ren2014cluscite}. These approaches 
satisfy the needs of a specific domain and are inapplicable to our studied problem. 
In our study, we need to jointly model the relationship between homogeneous nodes (paper) and the relationship between heterogeneous nodes (paper and dataset), also the text information of nodes (paper) for retrieving datasets based on the query specified as a list of papers as studied in \cite{altaf2019dataset}. However, our work is different from \cite{altaf2019dataset} that learn paper and dataset representation via Variational Autoencoders, as we  model the shared topics between paper and dataset nodes, to help us understand paper-paper citation and paper-dataset citation as well as to facilitate the query based dataset recommendation.

\begin{table}[t]
	\centering
	\renewcommand\tabcolsep{2.0pt}
	\caption{Comparison of STopic-Rec and other related works (Bi. - bipartite graph, Het. - heterogeneous graph)}
	\begin{tabular}{p{1.5cm}cccc}
		\hline 
		Method		& Bi./Het. & Content & Topic Modeling & Query\\ 
		\hline 
		\cite{wang2011collaborative, gopalan2014content, mcauley2013hidden, diao2014jointly,wang2015collaborative, kim2016convolutional, jin2018combining} 
		& Bi. & Item-Content & User-Topic \& Item-Topic  & No\\ 
		\hline 
		
		\cite{bao2014topicmf}		& Bi. & Item-Content & User-Topic-Item & No\\ 
		\hline 
		
		\cite{hu2015synthetic, hu2018collaborative, chen2014context}
		& Het. & Item-Content & User-Topic \& Item-Topic & No\\ 
		\hline 
		
		\cite{ma2009learning, ma2008sorec,Fan:2019:GNN,wu2019dual} & Het. & No & User-Topic \& Item-Topic  & No\\ 
		\hline 
		
		STopic-Rec		& Het. & User-Content & User-Topic-Item  & Yes\\ 
		\hline 
	\end{tabular} 
	\label{tab:comparison}
\end{table} 

\subsection{Topic Modeling in Recommendation}
Topic modeling has been widely used in content-based recommendation  \cite{pazzani2007content} 
which jointly models the ratings of users on items and attributes of items and users. Collaborative topic regression \cite{wang2011collaborative} was proposed to recommend scientific articles by combining traditional collaborative filtering and probabilistic topic modeling. 
McAuley et al. \cite{mcauley2013hidden} designed a transformation function to model the latent factors of ratings and reviews tightly. 
Similarly, Bao et al. \cite{bao2014topicmf} proposed a non-negative matrix factorization based method to bridge the factors of topics and ratings. Furthermore, Ling et al. \cite{ling2014ratings} adopted a mixture of Gaussian to replace matrix factorization to avoid the choice of the transformation function.
Recent works \cite{wang2015collaborative, kim2016convolutional, jin2018combining} applied Stacked Denoising Autoencoder \cite{vincent2010stacked}, CNN \cite{krizhevsky2012imagenet} and LSTM \cite{hochreiter1997long} incorporating with probabilistic matrix factorization to encode the review and improve the rating prediction accuracy.
Although these approaches investigated topic modeling in content-based recommendation, they neglect the useful social information between users. We summarize these approaches in Table \ref{tab:comparison} (the first two lines). They work on bipartite graphs with item content (without social links among users). They profile users and items in different topics, except \cite{bao2014topicmf} which applied topic modeling on the    review  content of a user-item pair. 
Another group of previous work in 
\cite{ma2009learning, ma2008sorec,Fan:2019:GNN,wu2019dual} investigated topic model on heterogeneous graphs (bipartite graphs + social links), but without item content. Similarly, user and item are profiled in separated topic distributions.
Recently, several works \cite{chen2014context, hu2015synthetic, hu2018collaborative} 
proposed hybrid recommendation methods which jointly model ratings, reviews and social information to learn \textit{different topics for users and items}. 
Though individual user topics and item topics have been widely leveraged for predicting user item rating, it is not appropriate to model separate topics in our scenario.
We aim  to model paper-paper relationship (social relationship), paper and dataset relationship (ratings), and paper-text information (reviews). Specifically, the paper text contents and dataset should share common topics distribution because the datasets used in paper and text content of paper both share the similar topic information of paper, unlike user and item in recommendation which are in different domains.  

\begin{table}[!t]
	\centering
	\caption{Notations}
	\begin{tabular}{cc}  
		\hline
		Symbol & Description\\
		\hline  
		$p \in \mathcal{V}_p$ & A paper \\
		$d \in \mathcal{V}_d$ & A dataset \\
		$w \in \mathcal{V}_w$ &  A word \\
		$\mathbf{v}$ & Embedding vector for a paper\\
		$K$ & Embedding dimension (number of topics)  \\
		$\theta$ & Topic distributions for text corpus $\mathcal{ T}$ \\
		$\phi^w$ & Word distributions for topics \\
		$\phi^d$ & Dataset distributions for topics \\
		$\bm{\psi}^w$ & Word embedding vector\\
		$\bm{\psi}^d$ & Dataset embedding vector\\
		$\kappa$ & Peakiness Parameter in transformation function\\
		$z_{p,j}^w$ & Topic for the $j$-th word in paper $p$\\
		$z_{p,i}^d$ & Topic for the $i$-th dataset in paper $p$\\
		$w_{p,j}$ & $j$-th word in paper $p$\\
		$d_{p,i}$ & $i$-th dataset in paper $p$\\
		$N_{p,w}$ & Total number of words in paper $p$\\
		$N_{p,d}$ & Total number of datasets in paper $p$\\
		\hline
	\end{tabular}
	\label{tab:Notations}
\end{table}

\section{Preliminaries}
In this section, we define several concepts to be used, and formally formulate the problem of query based dataset recommendation. Notations used throughout the paper are summarized in Table \ref{tab:Notations}. 
\newtheorem{definition}{Definition}
\begin{definition}
	\textbf{Dataset.} A dataset refers to  a set of data samples (e.g., images, documents, purchase records) which can be used for specific research purpose, and usually be cited in different forms in research papers.
\end{definition}
\begin{definition}
	\textbf{Dataset Recommendation.} Given a heterogeneous graph $\mathcal{G} = \left( \mathcal{V}, \mathcal{E}\right)$ consisting of three different kinds of nodes $\mathcal{V}$ (\textit{papers} $\mathcal{V}_p$, \textit{datasets} $\mathcal{V}_d$ and \textit{words} $\mathcal{V}_w$) and three different kinds of edges $\mathcal{E}$ (\textit{paper-cite-paper} $\mathcal{E}_p$, \textit{paper-use-dataset}  $\mathcal{E}_d$ and \textit{paper-contain-word} $\mathcal{E}_w$), and a query consisting of a set of few papers describing the user's research interest $\mathcal{Q} = \left\lbrace p_i \right\rbrace_{i=\left\lbrace 1,..., |\mathcal{Q}|\right\rbrace } \subseteq \mathcal{V}_p$, the goal is to recommend a set of top-$k$ datasets  $\left\lbrace d_i \right\rbrace_{i=\left\lbrace 1,..., k\right\rbrace} \subseteq \mathcal{V}_d$ that are ranked in order of relevance to the query $\mathcal{Q}$.
\end{definition}
\begin{definition}
	\textbf{Paper Citation Network.} A paper citation network is defined as a graph $\mathcal{G}_p = (\mathcal{V}_p, \mathcal{E}_p)$ which is a subgraph of $\mathcal{G}$. The adjacency matrix of graph $\mathcal{G}_p$ is noted as $\mathbf{A}^p \in \mathbb{R}^{|\mathcal{V}_p| \times |\mathcal{V}_p| }$.
\end{definition}

\section{Proposed Model}
Fig.2 shows the overall framework of our proposed dataset recommendation model  including four parts discussed below.
\begin{figure*}
	\centering
	\begin{tikzpicture}
	\tikzstyle{main}=[circle, minimum size = 10mm, thick, draw =black!80, node distance = 8mm]
	\tikzstyle{connect}=[-latex, thick]
	\tikzstyle{box}=[rectangle, draw=black!100]
	
	
	\node[main, fill = black!20] (A){$\mathrm{A}^{p}$};
	\node[main] (bdi) [above=of A] {$\phi^w$};
	\node[main] (adi) [above=of bdi] {$\phi^d$};
	
	\node[main, fill = black!20] (yij) [below =of A] {$y_{ij}$};
	
	\node[main] (spi) [left=of A] {$\mathbf{s}_{p_i}$};
	\node[main] (spj) [right=of A] {$\mathbf{s}_{p_j}$};
	\node[main] (thetai) [left = of spi]{$\theta_i$};
	\node[main] (thetaj) [right=of spj] {$\theta_j$};
	\node[main] (tpi) [left=of thetai] {$\mathbf{t}_{p_i}$};
	\node[main] (tpj) [right=of thetaj] {$\mathbf{t}_{p_j}$};
	
	\node[main] (vpi) [below=of thetai] {$\mathbf{v}_{p_i}$};
	\node[main] (vpj) [below=of thetaj] {$\mathbf{v}_{p_j}$};

	\node[main] (zpi) [above=of thetai] {$z^w_i$};
	\node[main] (zpj) [above=of thetaj] {$z^w_j$};
	\node[main] (zdi) [above=of zpi] {$z^d_i$};
	\node[main] (zdj) [above=of zpj] {$z^d_j$};

	\node[main, fill = black!20] (wpi) [above=of tpi] {$w_{p_i}$};
	\node[main, fill = black!20] (wpj) [above=of tpj] {$w_{p_j}$};
	\node[main, fill = black!20] (dpi) [above=of wpi] {$d_{p_i}$};
	\node[main, fill = black!20] (dpj) [above=of wpj] {$d_{p_j}$};

	\path (thetai) edge [connect] (zpi)
	(thetaj) edge [connect] (zpj)
	(wpi) edge [connect] (zpi)
	(wpj) edge [connect] (zpj)
	(dpi) edge [connect] (zdi)
	(dpj) edge [connect] (zdj)
	(vpi) edge [connect, dashed, color=red] (thetai)
	(vpj) edge [connect, dashed, color=red] (thetaj)
	(vpi) edge [connect, dashed, color=blue] (yij)
	(vpj) edge [connect, dashed, color=blue] (yij)
	(spi) edge [connect, dashed] (vpi)
	(spj) edge [connect, dashed] (vpj)
	(wpi) edge [connect, dashed, color=orange] (tpi)
	(wpj) edge [connect, dashed, color=orange] (tpj)
	(A) edge [connect, dashed, color=orange] (spi)
	(A) edge [connect, dashed, color=orange] (spj)
	(tpi) edge [connect, dashed, color=black] (vpi)
	(tpj) edge [connect, dashed, color=black] (vpj)
	(adi) edge [connect] (zdi)
	(adi) edge [connect] (zdj)
	(bdi) edge [connect] (zpi)
	(bdi) edge [connect] (zpj); 
	\draw [connect] (thetai) [out=135, in=235] to (zdi);
	\draw [connect] (thetaj) [out=45, in=305] to (zdj);

	
	
	\node[rectangle, inner sep=0mm, fit= (dpi) (zdi),label=above left:$D_{p_i}$, yshift=-2mm, xshift=-5mm] {};
	\node[rectangle, inner sep=3.4mm,draw=black!100, fit= (dpi) (zdi)] {};
	
	\node[rectangle, inner sep=0mm, fit= (dpj) (zdj),label=above left:$D_{p_j}$, yshift=-2mm, xshift=-5mm] {};
	\node[rectangle, inner sep=3.4mm,draw=black!100, fit= (dpj) (zdj)] {};
	
	\node[rectangle, inner sep=0mm, fit= (wpi) (zpi),label=above left:$N_{p_i}$, yshift=-2mm, xshift=-5mm] {};
	\node[rectangle, inner sep=3.4mm,draw=black!100, fit= (wpi) (zpi)] {};
	
	\node[rectangle, inner sep=0mm, fit= (wpj) (zpj),label=above left:$N_{p_j}$, yshift=-2mm, xshift=-5mm] {};
	\node[rectangle, inner sep=3.4mm,draw=black!100, fit= (wpj) (zpj)] {};
	
	\node[rectangle, inner sep=0mm, fit= (adi),label=above left:$K$, yshift=-2.5mm, xshift=2.5mm] {};
	\node[rectangle, inner sep=1.8mm,draw=black!100, fit= (adi)] {};
	
	\node[rectangle, inner sep=0mm, fit= (bdi),label=above left:$K$, yshift=-2.5mm, xshift=2.5mm] {};
	\node[rectangle, inner sep=1.8mm,draw=black!100, fit= (bdi)] {};
	
	\end{tikzpicture}
	\caption{Graphical representation of our proposed STopic-Rec model. Shaded nodes represent observed variables. Dashed orange arrows denote joint representation learning module; dashed black arrows denote concatenation operation; dashed red arrows denote topic-embedding fusion module; dashed blue arrows denote paper pairwise modeling module. In addition, black arrows present the dependencies in probabilistic topic modeling.}
\end{figure*}
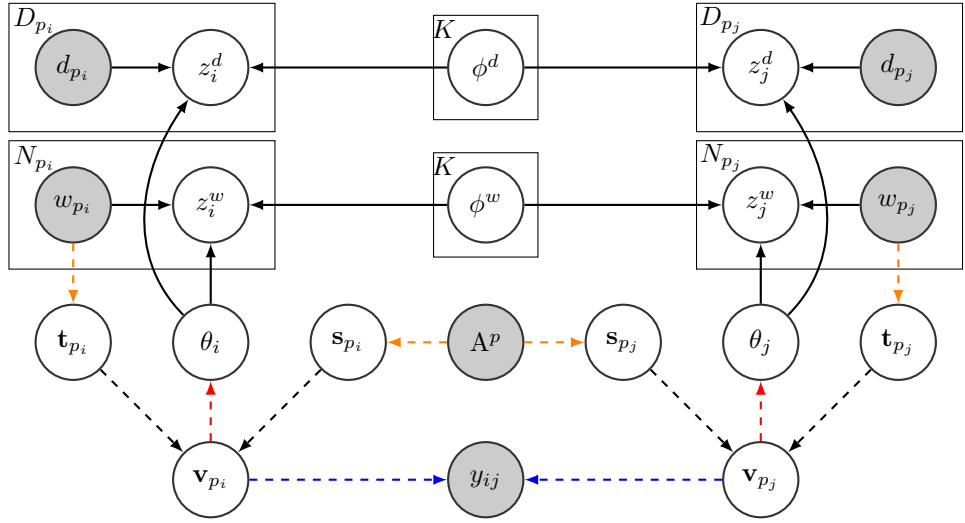

\subsection{Content and Structure based Node Representation}
To capture the important information associated with both the structure of paper citation network $\mathcal{G}_p$ and underlying associated content (e.g., text) of papers, we learn for each paper node a text-based embedding $\mathbf{t}$ and a structure-based embedding $\mathbf{s}$ separately and then simply concatenate both the embeddings to obtain a single representation $\mathbf{v}$ for a paper,
\begin{equation}
\mathbf{v} = [\mathbf{t}; \mathbf{s}]
\end{equation}
Next, we describe in detail the representation learning for text of papers and citation network structure respectively.

\subsubsection{\textbf{Encoding Global Text Information}}
Paper nodes are associated with words which are extracted from the content of papers such as title, abstract and keywords (if given).
To obtain the global text information of a paper, we apply Stacked Denoising Autoencoder \cite{vincent2010stacked} to encode the text information of each paper. An autoencoder is a feed-forward neural network that aims to reconstruct the given input by using a lower dimensional hidden layer \cite{vincent2010stacked}. Denoising autoencoder is an autoencoder with noise-corrupted input in order to avoid over-fitting and it guarantees that the learned high level representation should be stable and robust. Then we can construct the stacked denoising autoencoder by stacking denoising autoencoders to form a deep network by using the hidden representation of one as input to the next autoencoder. This method has been proved to lead to a better generalization \cite{vincent2010stacked}. Specifically, we define the one hidden layer denoising autoencoder as an example:
\begin{equation}
\mathbf{\tilde{x}} = \mathrm{Corrupted}(\mathbf{x})
\end{equation}
\begin{equation}
\mathbf{t} = \sigma(\mathbf{W}_1 \mathbf{\tilde{x}} + \mathbf{b_1})
\end{equation}
\begin{equation}
\mathbf{\hat{x}} = \sigma(\mathbf{W}_2 \mathbf{t} + \mathbf{b_2})
\end{equation}
where $\mathbf{x} = \mathbf{A}^w_{p,:}$ is the input text feature vector, $\mathbf{A}^w$ refers to the adjacency matrix that identifies the edges between paper nodes $\mathcal{ V}_p$ and word nodes $\mathcal{V}_w$, $\mathbf{\tilde{x}}$ is the corrupted input $\mathbf{x}$ by randomly setting some of elements to 0, $\mathbf{t}$ is the hidden vector which can be used as the embedding for text, then $ \mathbf{\hat{x}}$ is the reconstructed input feature vector, $\mathbf{W}_1$, $\mathbf{W}_2$, $\mathbf{b_1}$ and $\mathbf{b_2}$ are trainable parameters, and the activation function is set as LeakyReLU (Rectified Linear Unit) \cite{maas2013rectifier}.
The loss function is defined as follows for measuring the reconstruction loss:
\begin{equation}
\mathcal{L}_{t} = \sum_{i=1}^{|\mathcal{V}_p|}||\mathbf{x_i} - \mathbf{\hat{x_i}}||_2^2 + \lambda ||\mathbf{W}||_2^2
\end{equation}
where $\lambda$ is a regularization parameter, $|\mathcal{V}_p|$ denotes the size of text corpus and $\mathbf{W}$ is all trainable parameters. After training the stacked denoising autoencoders, we can obtain the embedding $\mathbf{t}$ for text of each paper.

\subsubsection{\textbf{Encoding Structural Information of Citation Network}}

To obtain the structural information of the given paper citation network $\mathcal{G}_p$, we leverage the widely used network embedding algorithm \cite{cui2018survey} to encode the structure information. The objective 
is to find a mapping function which converts each node in network to a low-dimensional latent representation, such that the nodes close in the topology of network should also have   small distances in latent feature space. This principle has been widely adopted in diverse network embedding algorithms \cite{cui2018survey}.  
Formally, 
following the idea in LINE \cite{tang2015line} (a popular network embedding algorithm),
the likelihood that paper $p_j$ is connected with paper $p_i$ is defined  as:
\begin{equation}
p(p_j|p_i) = \frac{\exp(\mathbf{s}_{p_j}\cdot \mathbf{s}_{p_i})}{\sum_{z \in P}\exp(\mathbf{s}_{p_i}\cdot \mathbf{s}_{p_z})}
\end{equation}
where $\mathbf{s}_{p_i}$ and $\mathbf{s}_{p_j}$ are the  embedding vector for paper $p_i$ and $p_j$, respectively. 
By using negative sampling \cite{mikolov2013distributed}, we have the objective function modeling neighborhood proximity as follows: 
\begin{equation}
\small
\begin{aligned}
\mathcal{L}_s &= \prod_{i=1}^{|\mathcal{V}_p|}\log p(\mathcal{N}_i|p_i) = \prod_{i=1}^{|\mathcal{V}_p|}\prod_{j \in \mathcal{N}_i}\log p(p_j|p_i) \\ &=
\prod_{i=1}^{|\mathcal{V}_p|}\prod_{j \in \mathcal{N}_i}(\log\sigma (\mathbf{s}_{p_i}^T\mathbf{s}_{p_j}) +\sum_{g=1}^{k}E_{p_x\sim F(p)}[\log \sigma(-\mathbf{s}_{p_i}^T\cdot \mathbf{s}_{p_x})] )
\end{aligned}
\end{equation}
where $\mathcal{N}_i$ is the neighbor set of node $p_i$,  $k$ is the number of negative samples, $\sigma$ denotes the sigmoid function and $\mathbf{s}_{p_x}$ is the embedding of a negative sample. $F(p)\propto d_p^{3/4}$ represents the distribution of papers, where $d_p$ is the out-degree of $p$. The function
$\mathcal{L}_s$ should be maximized to produce embedding vector $\mathbf{s}_{p_i}$ close to $\mathbf{s}_{p_j}$  for neighboring paper $p_i$ and $p_j$, and $\textbf{s}_{p_i}$ distant to $\mathbf{s}_{p_x}$ because $p_x$ is a negative sample. Meanwhile, in our work, we follow the neighborhood sampling strategy in Node2vec \cite{grover2016node2vec} to get paths in the network. From network embedding, we can obtain the embedding $\mathbf{s}$ for each paper.

\subsection{Probabilistic Topic Modeling}
The goal of the probabilistic topic modeling is to infer the word probability for a topic, and the dataset probability for a topic. Previous topic modeling based recommendation methods \cite{chen2014context, hu2015synthetic, hu2018collaborative} 
have demonstrated the effectiveness of using topic modeling to encode the text information. 
In our scenario, the dataset should share the same topic distribution with the word in text of paper, because the text and used dataset all are the content of the paper and they should present the similar topics with the research idea of the paper. Therefore, we grant the same topic distribution for both text and dataset. 

Following the notation in the past topic modeling work~\cite{blei2003latent}, we let $\mathbf{\theta}_p = \left\lbrace \theta_{p,k} \right\rbrace_{k=1}^K $ be the topic distribution for paper $p$, and $\sum_{k=1}^{K}\theta_{p,k} = 1$. We also let $\mathbf{\phi}^w = \left\lbrace \phi^w_{k} \right\rbrace_{k=1}^K $ be the word distribution over topics. $\phi^w_k = \left\lbrace \phi^w_{k, j}\right\rbrace_{j = 1}^{|\mathcal{V}_w|} $ is a multinomial distribution of words for topic $k$, while the probability of a word $j$ belonging to $k$, $\phi^w_{k, j} = P(j|k) > 0$, and $\sum_{j=1}^{|\mathcal{V}_w|} \phi^w_{k, j} = 1$; $|\mathcal{V}_w|$ is the size of the vocabulary $\mathcal{V}_w$. To ensure $\phi^{w}_k$ the word distribution for topic $k$ is a stochastic vector ($\sum_{j=1}^{|V|} \phi^w_{k, j} = 1$), we introduce an additional variable $\bm{\psi}^w$ and define
\begin{equation}
\phi^w_{k, j}=\frac{exp(\psi^w_{k,j})}{\sum_{j'} exp(\psi^w_{k,j'})}
\end{equation}

Similarly, we let $\mathbf{\phi}^d = \left\lbrace \phi^d_{k} \right\rbrace_{k=1}^K $ be the dataset distribution over topics. $\phi^d_k = \left\lbrace \phi^d_{k, i}\right\rbrace_{i = 1}^{|\mathcal{V}_d|} $ is a multinomial distribution of datasets for topic $k$, while the probability of a dataset $i$ belonging to $k$, $\phi^d_{k, i} = P(i|k) > 0$, and $\sum_{i=1}^{|\mathcal{V}_d|} \phi^d_{k, i} = 1$; $|\mathcal{V}_d|$ is the size of the set of dataset $\mathcal{V}_d$. To ensure $\phi^{d}_k$ the dataset distribution for topic $k$ is a stochastic vector ($\sum_{i=1}^{|\mathcal{V}_d|} \phi^d_{k, i} = 1$), we similarly introduce an additional variable $\bm{\psi}^d$ and define 
\begin{equation}
\phi^d_{k, i}=\frac{exp(\psi^d_{k,i})}{\sum_{i'} exp(\psi^d_{k,i'})}
\end{equation}
We can optimize $\bm{\psi}^d \in \mathbb{R}^{|\mathcal{V}_d|}$ and $\bm{\psi}^w \in \mathbb{R}^{|\mathcal{V}_w|}$ to determine $\phi^d  \in \triangle^{|\mathcal{V}_d|}$ and $\phi^w \in \triangle^{|\mathcal{V}_w|}$, here $\bm{\psi}^d$ and $\bm{\psi}^w$ act as the natural parameters for the multinomial distribution $\phi^d$ and $\phi^w$. In later discussion, we still use $\phi^w$ and $\phi^d$ for convenience, then use $\bm{\psi}^w$ and $\bm{\psi}^d$ in optimization.

Therefore, our model includes word distributions for each topic $\phi^w_k$, dataset distributions for each topic $\phi^d_k$, and topic distributions $\theta_p$ for each paper $p$, also the topic assignments  for each word $z^w_{p, j}$ and the topic assignments  for each dataset $z^d_{p,i}$. The joint distribution of a particular text corpus $\mathcal{T}$ and a set of dataset $\mathcal{D}$ is defined as follows (given the word distribution, dataset distribution, and topic assignments for each word and each dataset):
\begin{equation}
\begin{aligned}
&p(\mathcal{T}, \mathcal{D}|\theta, \phi^w, \phi^d, z^w, z^d) \\&= \prod_{p \in \mathcal{V}_p}  \prod_{j=1}^{N_{p, w}}\theta_{p, z^w_{{p}, j}}\phi^w_{z_{{p}, j}^w, w_{p, j}} \prod_{i=1}^{N_{p,d}} \theta_{p, z^d_{{p}, i}}\phi^d_{z^d_{{p}, i}, d_{p, i}}
\end{aligned}
\end{equation}
where $\theta_{p, z^w_{{p}, j}}$ and $\theta_{p, z^d_{{p}, i}}$ denote the likelihood of seeing particular topics for a given research paper, $\phi^w_{z^w_{{p}, j}, w_{p, j}}$ denotes the likelihood of seeing particular words for this topic,  similarly, $\phi^d_{z^d_{{p}, i}, d_{p, i}}$ refers to the likelihood of seeing particular datasets for this topic. Besides, $N_{p,w}$ is the number of words in paper $p$, and $N_{p,d}$ is the number of datasets in paper $p$.

\subsection{Topic-Embedding Fusion}
To fuse the joint representation  $\mathbf{v}_{p}$  learned from text and structure of a paper node into the topic generation process, we need a transformation function that allows arbitrary $\mathbf{v}_{p} \in \mathbb{R}^K$, while enforcing $\theta_{p} \in \triangle^K (i.e. \sum_k \theta_{p,k}=1)$. Also the transformation function should be monotonic, because it should preserve orderings so that the largest values of $\mathbf{v}_{p}$ should also be the largest values of $\theta_{p}$. Thereby, inspired by HFT model \cite{mcauley2013hidden}, we define the transformation function as follows:
\begin{equation}
\theta_p = \mathrm{Transformation}(\mathbf{v}_{p})
\end{equation}
Specifically, each element in $\theta_p$ is formulated as:
\begin{equation}
\theta_{p,k}=\frac{exp(\kappa v_{p,k})}{\sum_{k'}exp(\kappa v_{p,k'})}
\end{equation}
where the parameter $\kappa$ controls the `peakiness' of the transformation. As $\kappa \rightarrow \infty$, $\theta_{p}$ will approximate a unit vector that takes the value $1$ only for the largest index of $\mathbf{v}_{p}$; as $\kappa \rightarrow 0$, $\theta_{p}$ approximates a uniform distribution \cite{mcauley2013hidden}. Intuitively, a large $\kappa$ means that a paper only discusses the most \textit{important} topics, while a small $\kappa$ means that a paper discussed all topics evenly. In our work, we do not fit both $\mathbf{v}_{p}$ and $\theta_{p}$, because one defines the other (in practice we only fit $\mathbf{v}_{p}$).
This aligns hidden factors in embedding of a paper with hidden topics in text of the paper. 

\subsection{Paper Pairwise Modeling}
For network embedding, we desire that the connected papers in paper citation graph $\mathcal{G}_{p}$ share similar embeddings. In that sense, the more similar  $\mathbf{v}_{p_i}$ and $\mathbf{v}_{p_j}$ are, the higher  the $P(y_{p_i,p_j}=1|\mathbf{v}_{p_i}, \mathbf{v}_{p_j}, \eta)$ is. Since $\mathbf{v}_{p_i}$ and $\mathbf{v}_{p_j}$ are embedding vectors, their ``similarity" can be measured in terms of `cosine distance'. 
Here, we associate each edge $e_{ij} \in \mathcal{E}_p$ with a binary random variable denoted by $y_{p_i, p_j}$, with a value of 1 if the edge is present ($e_{ij} \in \mathcal{E}_p$) and 0 otherwise ($e_{ij} \notin \mathcal{E}_p$).
To transform this distance into a probability value, we define that:
\begin{equation}
\begin{aligned}
&P(y_{p_i,p_j}|\mathbf{\beta}^T,\mathbf{v}_{p_i}, \mathbf{v}_{p_j}) \\&= p^{y_{p_i,p_j}}(1-p)^{1-y_{p_i,p_j}} = \begin{cases}  
(1-p), \; \text{if} \, y_{p_i,p_j}=0 \\
p, \qquad \text{if} \, y_{p_i,p_j}=1,
\end{cases}
\end{aligned}
\label{equ:probabilityDistance}
\end{equation}
where $y_{p_i,p_j} \in \{0,1\}$ indicates if there is a citation link between $p_i$ and $p_j$ or not, and $\beta=[\beta_0,\beta_1]$ such that $\beta_0$ and $\beta_1$ are the bias and weights of neural logistic regression.

Then we define the likelihood when the citation link exists as follows:
\begin{equation}
P(y_{p_i,p_j}=1| \beta, \mathbf{v}_{p_i}, \mathbf{v}_{p_j})=  p^{y_{p_i,p_j}}= \sigma(\beta_0+\beta_1^T (\mathbf{v}_{p_i}\cdot\mathbf{v}_{p_j}))
\end{equation}
where $\sigma$ denotes the logistic function given as $\sigma(x)=\frac{1}{1+e^{-\beta_0+\beta_1\, x}}$, $\mathbf{v}_{p_i}$ and $\mathbf{v}_{p_j}$ are the $K$-dimensional embeddings of paper $p_i$ and $p_j$ and $\mathbf{v}_{p_i}\cdot\mathbf{v}_{p_j}$ is element-wise multiplication of embeddings for paper $i$ and paper $j$, respectively. The derivative of $\sigma(x)$ is $\sigma(x)(1-\sigma(x))$ following chain rule. 
For considering negative observations ($y_{p_i,p_j}=0$), we sample randomly for each paper from all non-existing links with other papers, by considering the equal number of negative paper pair links as the positive pair links.


\subsection{Optimization}
\label{sec:modelOptimization}

The objective of our model is to learn the optimal $\mathbf{V}$ for accurately modeling the connection between papers, and at the same time, obtain the most likely topics according to text and datasets of papers with the constraint of transformation function. From the description of previous sections, we can obtain the loss function as follows:
\begin{equation}
\begin{aligned}
\mathcal{L} &= \sum_{(p_i,p_j) \in\mathcal{ E}_p}\log p(y_{p_i,p_j}=1|\mathbf{v}_{p_i}, \mathbf{v}_{p_j}) \\ & +
\sum_{(p_i,p_j) \notin \mathcal{E}_p}\log p(y_{p_i,p_j}=0|\mathbf{v}_{p_i}, \mathbf{v}_{p_j}) \\ & -
\lambda  \sum_{p}^{|\mathcal{V}_p|} \sum_{j=1}^{N_{p, w}} \log (\theta_{p, z^w_{p,j}}\phi^w_{z^w_{{p}, j}, w_{p, j}}) 
\\& - \lambda \sum_{p}^{|\mathcal{V}_p|}\sum_{i=1}^{N_{p, d}} \log (\theta_{p, z^d_{{p}, i}}\phi^d_{z^d_{{p}, i}, d_{p, i}}) \\ & + 
\lambda_p \sum_{p=1}^{|\mathcal{V}_p|} \mathbf{v}_{p} \mathbf{v}_{p}^T+\lambda_d \sum_{i=1}^{|\mathcal{V}_d|} \bm{\psi}^d_{i} {\bm{\psi}^d_{i}}^T+\lambda_v \sum_{i=1}^{|\mathcal{V}_w|} \bm{\psi}^w_{i} {\bm{\psi}^w_{i}}^T 
\\& +\lambda_{W} \sum_{k=1}^{N_k} ||\mathcal{W}||^2_F
\end{aligned}
\label{equ:lossFunction}
\end{equation}
where $(p_i,p_j)$ denotes all the paper pairs, $w_{p,j}$ is the $j$-th word in paper $p$, $d_{p,i}$ is the $i$-th dataset used in paper $p$, $z^w_{p,j}$ is the word's corresponding topic, and $z^d_{p,i}$ is the dataset's corresponding topic, $\lambda_{p}$, $\lambda_d$ and $\lambda_v$ are corresponding regularization parameters. 
$\mathcal{W}$ is the set of weight parameters in stacked denoising autoencoder and network embedding model, we first pretrain them, then jointly fine-tuning the parameters with topic modeling module and pairwise modeling module. $N_k$ is the number of weights in $\mathcal{ W}$. 
Moreover, $\lambda$ is a parameter that balances the performance of link prediction between papers, and topic modeling. Last four terms are all regularization for embedding vectors and weight vectors for auto-encoder respectively.

Our goal is to simultaneously optimize the parameters associated with  papers $\Theta=\{\mathbf{V}\}$,  the parameters associated with topics $\Phi=\{\theta, \phi^w, \phi^d\}$, the parameters associated with paper structure embedding and text embedding, $\Omega=\{\mathcal{W}\}$.
That is, given a heterogeneous graph $\mathcal{G}$, our objective is to find:
\begin{equation}
\mathrm{argmin}_{\Theta,\Phi,\Omega, \kappa, z^w,z^d} \mathcal{L}(\mathcal{G}|\Theta,\Phi,\Omega,\kappa, z^w,z^d)
\end{equation}
The optimization process is shown as follows:

\begin{itemize}
	\item[(1)] Solve the objective by fixing $(z^{w})^t$, $(z^{d})^t$ and $\Omega^t$:
	\begin{equation}
	\mathrm{argmin}_{\Theta,\Phi ,\kappa} \mathcal{L} (\mathcal{G}| \Theta,\Phi,(z^{w})^t, (z^{d})^t,\kappa,\Omega^t)
	\end{equation}
	to update $\Theta^{t+1}$, $\Phi^{t+1}$ and $\kappa^{t+1}$.
	
	\item[(2)]
	\begin{itemize}
		\item[(a)] Update $\Omega^{t+1}$ with fixing $\Theta$.
		
		\item[(b)] Sample $(z^{w}_{p,j})^{t+1}$ with probability
		\begin{equation}
		p((z^{w}_{p,j})^{t+1}=k)=(\phi^{w}_{k,w_{p, j}})^{t+1}
		\end{equation}
		
		\item[(c)] Sample $(z^{d}_{p,i})^{t+1}$ with probability
		\begin{equation}
		p((z^{d}_{p,i})^{t+1}=k)=(\phi^{d}_{k,d_{p, i}})^{t+1}
		\end{equation}

	\end{itemize}
	
\end{itemize}

In step (1), we fix $z^w$, $z^d$ and $\Omega$ to update remaining terms $\Theta$, $\Phi$, $\kappa$ by using L-BFGS algorithm \cite{nocedal1980updating}. In  step (2), we fix $\Theta$, $\Phi$, $\kappa$ to update the parameters $\Omega$ of joint representation learning from text and structure information, and parameter $z^w$, $z^d$ of probabilistic topic model. We can update $\Omega$ and $z^w$, $z^d$ separately, because both modules are independent once the $\Theta$ is fixed. In step 2(a), we update $\Omega$ by using back propagation algorithm. Then, in step 2(b), we iterate through all papers and each word within to update $z^w_{p, j}$ via Gibbs sampling. Similarly, in step 2(c), we iterate through all papers and each dataset within to update $z^d_{p, i}$ via Gibbs Sampling. We repeat step (1) and (2) until convergence. For computing the gradients of loss function w.r.t. different parameters, mathematical derivation  is given in Appendix. 
\section{Experiments}

In this section, we 
answer the following research questions:
\begin{itemize}
	\item (\textbf{RQ1}) Does our model perform better for dataset recommendation than the state-of-the-art recommendation methods?
	\item (\textbf{RQ2}) How do different hyper-parameters in our model influence the dataset recommendation?
	\item (\textbf{RQ3}) What is the effect of different model components in our framework for dataset recommendation?
	\item (\textbf{RQ4}) Are our generated dataset profiles reasonable?
\end{itemize}

\subsection{Experiment Design}

\subsubsection{\textbf{Datasets}}
We use an academic network data from Delve system \cite{akujuobi2017delve} and remove all isolated papers that have no links to any other papers or datasets. After preprocessing, the dataset includes 8,519 papers and 5,610 datasets. Each paper has a text file that contains title, abstract and keywords (if given). We divide the whole dataset into training data and query data.\\ \textbf{Training data}: we use 8,093 papers published from 1991 to 2014, and 5,610 datasets leveraged in these papers to train our proposed model. The input network has 360K citations among papers and 14K bipartite links between papers and datasets. \\
\textbf{Query data}: we use 426 papers published from 2015 to 2016 which are not in the training data as the test data for evaluating dataset recommendations. Then we select a set of papers from the reference list of one paper in test data to form a query for the paper if they are available in training data. After that, we obtain 426 queries with ground-truth datasets that were selected by paper's authors in the corresponding paper.

\subsubsection{\textbf{Baselines}} 

For baselines, we consider 17 different algorithms spanning 5 different categories as discussed below. 

\begin{enumerate}
	\item[(a)]{\textbf{Naive Techniques}
		\begin{itemize}[leftmargin=-0.1cm]
			\item{\textbf{Naive Retrieval}: A naive way to retrieve relevant datasets for the query specifying relevant papers is to return all the first order neighboring datasets for given papers and rank these datasets by their frequency of appearance in the retrieved results.}
			\item{\textbf{Advanced Naive Retrieval}:  An advanced approach would be to retrieve $k$-order neighboring datasets and then order in descending manner w.r.t. to their frequency in the retrieved results. Here we consider $k=2$, i.e. up to 2-hop neighboring datasets for specified papers in the query.}
		\end{itemize}
	}
	
	\item[(b)]{\textbf{Matrix Factorization Techniques}: iMF \cite{hu2008collaborative}, BPR \cite{rendle2009bpr}.
	}

	\item[(c)]{\textbf{Network Representation Learning (NRL) Techniques}: Deepwalk \cite{perozzi2014deepwalk}, Node2vec \cite{grover2016node2vec}, Metapath2vec \cite{dong2017metapath2vec}.
	}
	
	\item[(d)]{\textbf{Network and Content Representation Learning Techniques}: Text-associated Deepwalk (TADW) \cite{yang2015network}, NRL+ LSI \cite{hofmann2017probabilistic}, NRL+ Doc2vec \cite{le2014distributed}, GCN \cite{kipf2017semi}, HVGAE \cite{altaf2019dataset}.
		
	}
	
	\item[(e)]{\textbf{Topic Modeling based Techniques}: HFT \cite{mcauley2013hidden}, CDL \cite{wang2015collaborative}, CVAE-CF \cite{li2017collaborative}, JMARS \cite{diao2014jointly}, CTFP \cite{gopalan2014content}.
		%
	}
\end{enumerate}

\begin{table*}[t]
	\centering
	\LARGE
	\caption{Ranking performance comparison (the best results of baselines are marked as * along with underline). The results in bold are our method. $\mathbf{A}^{p}$, $\mathbf{A}^{d}$ and $\mathbf{A}^{w}$ denote paper-paper links, paper-dataset links sand paper-word links, respectively.}    
	\label{tab:rankingDatasets}
	\footnotesize
	\begin{tabular}{c|c|c|ccccl}
		\hline
		Information Source & Algorithms & Dim. & Pre@5 & Recall@5 & NDCG@5 & MRR@5 & AUC@5  \\\hline
		$\mathbf{A}^{d} $
		&Naive Retrieval & NA 	&0.0761&	0.3803	&0.3384&	0.3264	&0.6986
		\\
		$\mathbf{A}^{d} + \mathbf{A}^{p}$    & Advanced Naive Retrieval & NA &	0.1179&	0.5896&	0.4659&	0.4277&	0.8293
		\\
		\hline
		
		\multirow{2}{*}{ $\mathbf{A}^{d} $}
		&iMF& 256	&0.0793&	0.3967	&0.3659&	0.3573	&0.7041
		\\
		&BPR&128&	0.1214&	0.6072&	0.4598&	0.4133&	0.8999
		\\
		\hline
		\multirow{6}{*}{$\mathbf{A}^{d} + \mathbf{A}^{p}$ }
		&Deepwalk &128&0&0&0&0&0.5497\\
		&Node2vec &128&0&0&0&0&0.5962\\
		&Metapath2vec&128 & $0.0033$& 	$0.0164$& 	$0.0085$& 	$0.0060$& 	$0.7913$\\    
		&Deepwalk+BPR&128	&0.1714&	0.8568&	0.5912	& 0.5118	&0.9722\\
		&Node2vec+BPR &128	&0.1742	&0.8709	&0.5870 &	0.4914	&0.9765\\
		&Metapath2vec+BPR & 64	&0.1476	&0.7379& 0.5377	&0.4626 &	0.9673
		\\
		\hline 
		\multirow{5}{*}{$\mathbf{A}^{d} + \mathbf{A}^{w} $ }  
		&HFT& 16 &0.1632&0.8358&0.4931&0.4402& 0.9731
		\\
		&CDL& 16 &\underline{0.1814*}	&\underline{0.9092*}	&0.5523	&0.4932	&0.9617
		\\
		&CVAE-CF & 16 & 0.1213 & 0.6066& 0.3922 & 0.3224 & 0.9678
		\\
		& CTPF&16  &$0.053$& 	$0.4640$& 	$0.0850$& 	$0.0600$& 	$0.7064$
		\\   
		& JMARS&16&  0.1197	&0.5979	&0.3927	&0.3230	&0.9476
		\\   
		\hline 
		\multirow{9}{*}{$\mathbf{A}^{d} + \mathbf{A}^{p} + \mathbf{A}^{w} $ }  
		&DeepWalk+LSI+BPR& 256&	0.1700	&0.8498&	0.5978&	0.5163&	0.9720
		\\ 
		&DeepWalk+Doc2vec+BPR & 256&	0.1751	&0.8754&	0.6155	&0.5284&	0.9741
		\\
		&Node2vec+LSI+BPR& 256& 	0.1566& 	0.7829& 	0.5816	& 0.4910	& 0.9775
		\\ 
		&Node2vec+Doc2vec+BPR & 256	&0.1735	&0.8674&	0.6289	&0.5382&	0.9745
		\\
		&MetaPath2vec+LSI+BPR& 256	&0.1340	&0.6698	&0.4978 &	0.4433	&0.9605
		\\
		&MetaPath2vec+Doc2vec+BPR& 256	&0.1513	&0.7567	&0.5512	&0.4862	&0.9699
		\\
		& TADW +BPR &256	&0.1536	&0.7680 &0.6473&0.6087&	0.9607\\
		& GCN & 16 & 0.1134 &0.5586&0.3786&0.3023&0.9329\\
		& HVGAE & 16& \underline{0.1845*} & \underline{0.9225*} & \underline{0.7153*} & \underline{0.6536*} & \underline{0.9876*}\\
		& \textbf{STopic-Rec} & 16   &\textbf{0.1923}&\textbf{0.9408}&\textbf{0.7561}&\textbf{0.6929}&\textbf{0.9894}
		\\
		\hline
		
	\end{tabular}
\end{table*}

\subsubsection{\textbf{Experimental Settings}}

To generate recommendation score for each query consisting of a set of papers, we first generate individual dataset recommendation score for each paper in the query by taking dot product between the representation vector of paper and the representation vector of dataset, and then aggregate the query preferences by averaging the dataset recommendation scores for all papers in the query. Then, we rank datasets for the given query by sorting datasets according to their recommendation scores in descending order. The higher the score value, the more relevant dataset is for the given query:
\begin{equation}
o(Q,d)= \frac{1}{|Q|} \sum_{p \in Q} \mathbf{v}_{p} \cdot \bm{\psi}^d   \qquad  \forall  d \in \mathcal{V}_d 
\end{equation}

We evaluate the performance of our proposed method STopic-Rec on prediction of top-$k$ recommended datasets by employing metrics  $precision@k$, $recall@k$, $ndcg@k$, $mrr@k$ and $auc@k$, where $k$ is the number of top datasets of interest, and 
higher values indicate better prediction performance.

To evaluate dataset recommendation performance, we used implementations for above baselines provided by their authors, fine-tuned the model settings and obtained the best performance on dataset recommendation task. For fair comparison with the baselines, we optimize and set the embedding dimension $K$ to the value that gives the best performance of each baseline for dataset recommendation task. 
For learning network based representation  learning (NRL) using Deepwalk, Node2vec, Metapath2vec, we use default parameters as shared in the original papers,  walk length as $80$, number of walks per node as $10$, and neighboring nodes in window size of $5$, as NRL approaches perform best with these parameter settings. Further, we learn content based representation using Doc2vec and LSI for text of paper that includes the title, abstract, and keywords (if given).  Since the network embedding is in the unsupervised setting and not optimized for dataset recommendation task, we use BPR \cite{rendle2009bpr} as the objective function to fine-tune the performance of Deepwalk, Node2vec, Metapath2vec and TADW w.r.t. pairwise ranking of papers and datasets. To generate metapaths for dataset recommendation, we define P-P, P-D-P, D-P-D as metapaths such that P-P refers to two papers  connected by a citation link, and P-D-P indicates two papers which share the same dataset, D-P-D refers to two datasets that are used for experimental evaluation in the same paper.
For topic modeling based collaborative filtering techniques, we learn topics and embedding from text of paper, and paper-dataset links only, then train for 50 epochs with default parameter settings that can be fine-tuned for our recommendation task. 
For our model, we set topic modeling regularization to $\lambda=0.5$, paper regularization  $\lambda_p=0.1$, dataset  regularization $\lambda_d=0.01$ and words regularization  $\lambda_v=0.1$,  regularization weights of stacked denoising autoencoder as $\lambda_w=0.01$, learning rate $lr=0.01$ and batch size is set to $64$ samples per batch. We first pre-train the embeddings for text in stacked de-noising auto-encoder framework and pre-train the network embedding using Node2vec, and then feed both embeddings separately to our STopic-Rec model for joint modeling and fine-tuning of papers, datasets, and word embeddings, training for $150$ epochs by fitting all parameters using LBFGS optimizer.

\subsection{Dataset Recommendation Performance Evaluation (RQ1)}
Table \ref{tab:rankingDatasets} shows the performance comparison of our proposed STopic-Rec model with all baseline methods. 
For the competing methods, we observe that
\begin{itemize}[leftmargin=0.3cm]
	\item{Among naive techniques, Advanced Naive Retrieval has better performance, since it considers $k$-hop papers  for each paper in a query to aggregate preferences for datasets based on their frequency unlike basic Naive Retrieval that only considers the first-order paper-dataset relations to find relevant datasets for the given query. However, these naive techniques fail to consider the semantics of papers and datasets, also have quite poor performance compared to our model.}
	\item{Matrix factorization techniques only consider paper-dataset links to model their relationship, and fail to consider the semantics, hence result in poor performance.}
	\item{NRL techniques only leverage relationship between homogeneous nodes (paper) and heterogeneous nodes (paper and dataset). Among these, Node2vec+BPR and Metapath2vec+BPR are competitive to each other but are not good competitors for our model.}
	\item{Among topic modeling based collaborative filtering techniques that model the unstructured text of papers, and links between paper and datasets such as HFT, JMARS, CDL, CVAE-CF,  and CTPF, we observe that CDL is a strong baseline for our model followed by HFT.  This is because it designs a bayesian model to jointly model deep representation learning for content information via auto-encoder and collaborative filtering and considers sparseness of rating matrix. Unlike CDL, other topic modeling based collaborative filtering techniques such as JMARS, CTPF and CVAE-CF show poor performance on sparse ratings and implicit feedback data. }
	\item{Finally, we consider both homogeneous links (between paper nodes) and heterogeneous links (papers and datasets) together with the content of papers to learn semantics of papers and words. Our model specifically learns profile for paper-topics and words-topic, and datasets-topic.  Among these baselines, HVGAE performs better but it fails to leverage topic modeling of attributed nodes, which is an important aspect to explain the common context between paper and paper as well as paper and dataset that have citation links in between. 
	}
	\end{itemize}

	\subsection{Parameter Analysis (RQ2)}
	\subsubsection{Influence of the embedding size K}
\begin{figure}[!t] 
		\centering
		\subfigure[Varying K] {\includegraphics[scale=0.239]{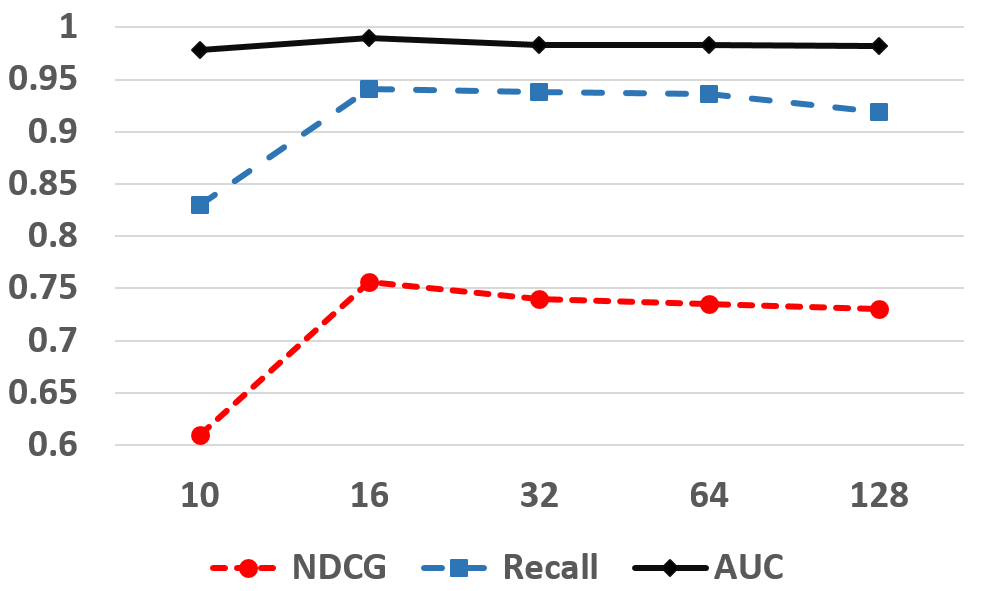}}
		\subfigure[Varying $\lambda$] {\includegraphics[scale=0.239]{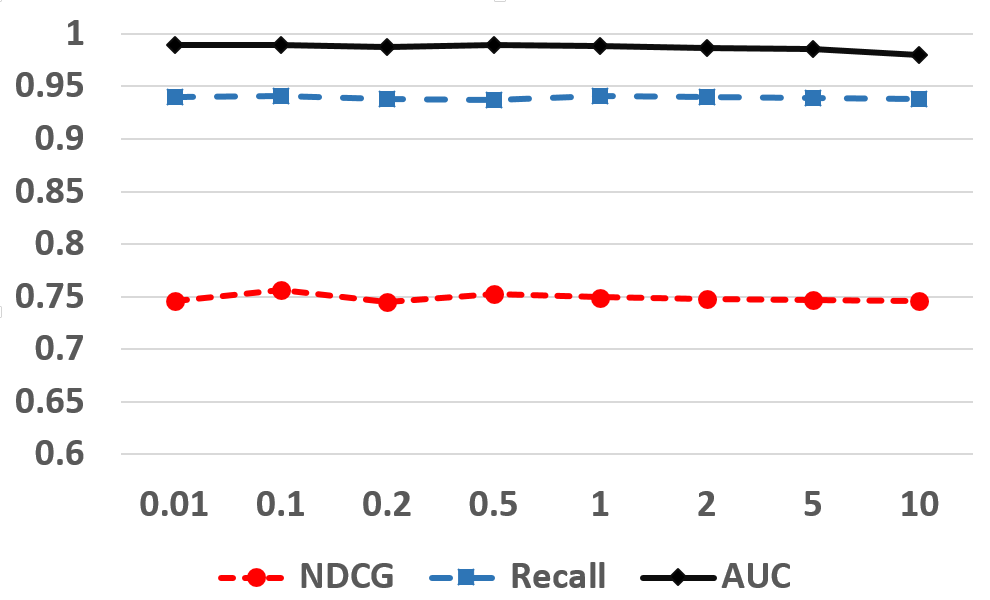}}
		\caption{Dataset Recommendation performance by varying (a) the number of latent topics and embedding size $K$,  and (b) topic modeling regularization parameters ($\lambda$).}
		\label{fig:parameterSensitivityAnalysis}
	\end{figure}

	First, we fix the regularization parameters to their default values of $\lambda_p=0.1$, $\lambda_d=0.01$ and $\lambda_v=0.1$ respectively, and vary the number of latent topics to be $10$, $16$, $32$, $64$ and $128$ respectively. As shown in Figure \ref{fig:parameterSensitivityAnalysis} (a), we observe that the performance of our model increases as we increase the number of latent topics up to $16$ and then remains relatively stable. Unlike generative models (e.g., LDA), we find that our model requires very few latent topics to exhibit the best performance. This is because one paper is mainly relevant to a few latent topics only.
	
	\subsubsection{Influence of the topic modeling regularization parameter $\lambda$ }
	Next, we fix the numberr of latent topics and embedding dimension to $16$, and vary topic modeling regularization parameters $\lambda$ for values of 
	$\{0.01,0.1, 0.2, 0.5, 1, 2, 5, 10\}$. As we can see from Figure \ref{fig:parameterSensitivityAnalysis} (b), the performance is relatively stable when $\lambda>=0.1$ for NDCG and AUC and $\lambda>=0.5$ for recall measure. This shows that the performance of our model is not sensitive to values of $\lambda$.

	
	
	\begin{figure}[!t] 
		\centering
		\includegraphics[scale=0.27]{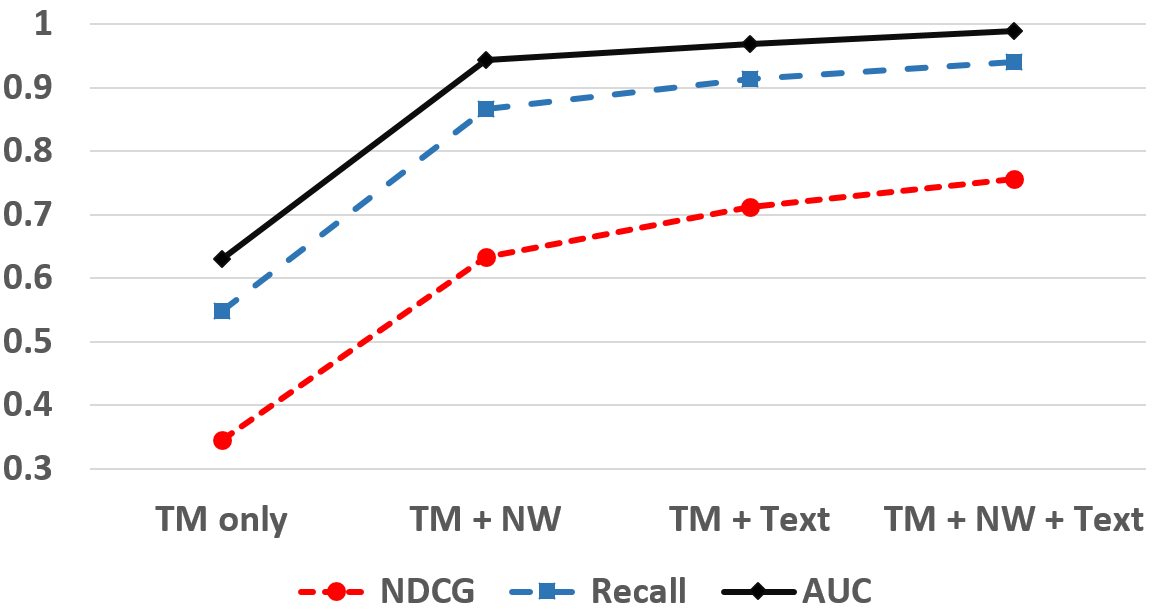}
		\small
		\caption{Ablation Study of STopic-Rec on different components.} 
		\label{fig:modelAblationStudy}
	\end{figure}

	\subsection{Model Ablation Study (RQ3)}
	We perform ablation study of our proposed model STopic-Rec by removing some features of the model and observe how that affects the performance. Our first ablated model named \textbf{TM} considers topic modeling along with the paper-paper citation links. Secondly, we consider topic modeling combined with network embedding (embedding of each papers using Node2vec) and pairwise link modeling between papers approach \textbf{TM+NW}. Third, we consider topic modeling combined with text modeling of papers along with pairwise link modeling of papers \textbf{TM+Text}. Finally, we consider all the components of our model \textbf{TM+NW+Text} that incorporates the network embedding, and text embedding of each paper, besides fusion with topic modeling and citation link modeling via logistic regression.
	Figure \ref{fig:modelAblationStudy} displays 
	\textbf{TM+NW} and \textbf{TM+Text} models perform comparatively better than \textbf{TM} demonstrating that incorporating the structure or text properties of each paper contribute well to the overall model. Finally, we observe that the combined model which couples topic modeling with joint text and network embedding   exhibits better prediction performance among all other models, demonstrating the need for joint modeling of all components together.
	
	\begin{table}[t]
		\centering
		\renewcommand\tabcolsep{2.0pt}
		\footnotesize
		\caption{CDL: Top five words from top $K$=4 topics for our dataset for the given query}
		\label{tab:top5words_Ktopics_cdl}
		\begin{tabular}{ccccl}
			\hline
			Topic 1&Topic 2&Topic 3&Topic 4 \\
			\hline
			\textbf{vector}&\textbf{cluster}&benchmark&management\\
			\textbf{linear}&attribute&\textbf{support}&retrieval\\
			simple&label&\textbf{machine}&optimal\\
			social&local&limit&best\\
			\textbf{class}&information&distribution&methods\\
			\hline
		\end{tabular}
	\end{table}

	\begin{table}[t]
		\centering
		\renewcommand\tabcolsep{2.0pt}
		\footnotesize
		\caption{Our Method: Top five words from top $K$=4 topics for our dataset for the given query}
		\label{tab:top5words_Ktopics_ours}
		\begin{tabular}{ccccl}
			\hline
			Topic 1&Topic 2&Topic 3&Topic 4 \\
			\hline
			\textbf{vector}&\textbf{linear}&\textbf{cluster}&discover\\
			\textbf{cluster}&\textbf{support}&\textbf{attributes}&estimation\\
			analyze&machine&local&analyze\\
			\textbf{pattern}&framework&patterns&empirical\\
			\textbf{linear}&simple&distribution&investigate\\
			\hline
		\end{tabular}
	\end{table}
	
	\begin{table}
		\centering
		\renewcommand\tabcolsep{2.0pt}
		\footnotesize
		\caption{Our Method: Top five datasets from top K=4 topics for Delve dataset for the given query}
		\label{tab:top5datasets_Ktopics_ours}
		\begin{tabular}{cccccl}
			\hline
			Topic 1&Topic 2&Topic 3&Topic 4  \\
			\hline
			d3&\textbf{d23}&d3&d29 \\
			\textbf{d23}&d932&d29&d5\\
			d8932&\textbf{d88}&d5&d11\\
			d723&d57&d4&d6\\
			\textbf{d292}&d11&d10&d4\\
			\hline
		\end{tabular}
	\end{table}
	\subsection{Case Study (RQ4)}
	To gain better insight into our model STopic-Rec, we perform  case study to compare our model with competitive topic modeling baseline \textbf{CDL}. Here we consider top $k$=$4$ topics for each query by averaging the preference scores of paper topics for all papers in the query, and then ranking topics in descending order for each query i.e., $\theta_{q,k}=\frac{1}{|Q|}\sum_p \theta_{p,k} \, \forall p \in Q,  k \in K$.
	We take an example query $Q=\{p51,p1938\}$ including two papers that are both related to \textit{Rule Extraction from Support Vector Machines (SVM)}.  The true datasets used for experimental evaluation in these papers are  $\{Glass04v5(d23),Led7digit(d292),Sonar(d88)\}$. 
	Both CDL and our model STopic-Rec capture topics information extracted both from auto-encoder framework, and topic modeling resulting in better word clustering and dataset clustering for each topic. Additionally, our model also captures dataset profiles for each topic.  In Table \ref{tab:top5words_Ktopics_cdl} and \ref{tab:top5words_Ktopics_ours}, we show the top 4 topics and top 5 words for each topic as a result of our query. 
	In Table \ref{tab:top5datasets_Ktopics_ours}, we observe that our model effectively clusters datasets which are related to similar topics e.g.$ \{d3, d29, d5,d4,d10\}$ are the most effectively used datasets for clustering, while $\{d23,d292,d88\}$ are most commonly used for evaluation of support vector machine based papers.

	\section{Conclusion}
	We propose a model for recommending scientific datasets as the answer to a query specified as a list of  papers. The relevance of dataset candidates to the query papers is evaluated by profiling them on a set of latent topics, which is shared between datasets and text in 
	paper(s). We learn representation of paper and dataset nodes in an attributed heterogeneous graph both from the node attributes and its structural properties. 
	Besides, we design a topic model to learn the shared topics of datasets and papers. Finally, we employ a transformation function to couple the relationship between text and structure based embedding  of nodes; and the inherent topics of each paper and dataset node.  
	Further, our algorithm results in interpretable profiles for papers and datasets by learning topics discussed in text from paper embeddings and estimating topic-words distribution and topic-datasets distribution. Such profiles are helpful in real-world interpretable recommendation system.

	\bibliographystyle{IEEEtran}
	\bibliography{IEEEabrv}
	
	\appendix
	\section{Gradient Derivation}
	Gradients of different parameters w.r.t. loss function are computed as follows:
	\begin{equation}
	\begin{aligned}
	\frac{\partial \mathcal{L}}{\partial \mathbf{v}_{p}}&= -\lambda \kappa \Big(  \sum_{k=1}^K (n_{p,k} - N_{p,w}\frac{exp(\kappa v_{p,k})}{\sum_k exp(\kappa v_{p,k})} ) \\& 
	+ \sum_{k=1}^K (d_{p,k} - N_{p,d}\frac{exp(\kappa v_{p,k})}{\sum_k exp(\kappa v_{p,k})} )  \Big)  
	+2 \lambda_p v_{p}  
	\end{aligned}
	\end{equation}
	
	\begin{equation}
	\frac{\partial \mathcal{L}}{\partial \kappa}= -\lambda \sum_{p=1}^{|\mathcal{V}_p|} \sum_{k=1}^K v_{p,k} \left(  n_{p,k}-N_{p,w} \frac{exp(\kappa v_{p,k})}{\sum_{k'}exp(\kappa v_{p,k'})}  \right)
	\end{equation}
	
	\begin{equation}
	\frac{\partial \mathcal{L}}{\partial \bm{\psi}^w}= -\lambda \sum_{j=1}^{|\mathcal{V}_w|} \sum_{k=1}^K \left( n_{k,j} - N_{j,k} \frac{exp(\psi^w_{k,j})}{\sum_{j'} exp(\psi^w_{k,j'})} \right)
	\end{equation}
	
	\begin{equation}
	\frac{\partial \mathcal{L}}{\partial \bm{\psi}^d}=-\lambda \sum_{i=1}^{|\mathcal{V}_d|} \sum_{k=1}^K \left( n_{k,i} - N_{i,k} \frac{exp(\psi^d_{k,i})}{\sum_{i'} exp(\psi^d_{k,i'})} \right)
	\end{equation}
	
	
	\begin{equation}
	\frac{\partial \mathcal{L}}{\partial{\beta_1}}=\frac{1}{|\mathcal{E}|}[(\sigma(\beta^T (\mathbf{v}_{p_i}\cdot\mathbf{v}_{p_j}))-y_{p_i,p_j})\cdot(\mathbf{v}_{p_i}\cdot\mathbf{v}_{p_j})]
	\end{equation}
	\begin{equation}
	\frac{\partial \mathcal{L}}{\partial{\beta_0}}=\frac{1}{|\mathcal{E}|} \sum_{\substack{(p_i,p_j) \in \mathcal{E}_p \\ (p_i,p_j) \notin \mathcal{E}_p}}[\sigma(\beta^T (\mathbf{v}_{p_i}\cdot\mathbf{v}_{p_j}))-y_{p_i,p_j}]
	\end{equation}
	
	where $\beta_1$, $\beta_0$ are the weights and bias of logistic regression, $y_{p_i,p_j}$ indicates if there is a citation link between $p_i$ and $p_j$, and $|\mathcal{E}|$ is the total number of positive and negative pairs for paper i and paper j such that $i\neq j$. $n_{k,j}$ denotes the number of times that word $j$ occurs in topic $k$; $n_{k,i}$ denotes the number of times that dataset $i$ occurs in topic $k$; $n_{p,k}$
	is the number of times when topic $k$ is selected for words in the paper $p$, and $d_{p, k}$ is number of times when topic $k$ is selected for datasets in paper $p$.
	Moreover, $N_{j, k}$ is the number of words in topic $k$; $N_{i, k}$ is the number of datasets in topic $k$; $N_{p,w}$ is the total number of words in paper $p$ and $N_{p,d}$ is the total number of datasets in paper $p$.

\end{document}